\documentclass[a4paper,11pt]{article}
\usepackage{amsmath}

\usepackage{aaskaiid}
\usepackage{aas_macros}%
\usepackage[utf8]{inputenc}
\hypersetup{
    colorlinks=true,
    linkcolor=red,
    filecolor=magenta,      
    urlcolor=magenta,
    citecolor=blue,
}
\usepackage{multirow}
\usepackage{tabularx} 
\usepackage{adjustbox} 
\usepackage{pdflscape} 
\usepackage{graphicx} 
\usepackage{lscape} 
\usepackage{longtable} 
\usepackage{booktabs} 
\usepackage{caption} 
\usepackage{adjustbox}
\usepackage{threeparttable}
\usepackage[utf8]{inputenc}
\usepackage{orcidlink}

\DeclareUnicodeCharacter{2033}{\textquotedbl}
\title{Understanding Coronal `Seeing' using High Angular Resolution Radio Observations of Active Solar Emissions}
\ShortTitle{Understanding Coronal `Seeing'}

\author[1]{Soham Dey\orcidlink{0009-0006-3517-2031}}
\ShortName{S. Dey et al.} % shortened name list for header 
\author[1]{Deepan Patra \orcidlink{0009-0003-8133-6621}}
\author[1]{Divya Oberoi \orcidlink{0000-0002-4768-9058}}
\author[2]{Anshu Kumari\orcidlink{0000-0001-5742-9033}}
\author[3,4]{Atul Mohan\orcidlink{0000-0002-1571-7931}}
\author[1]{Surajit Mondal \orcidlink{0000-0002-2325-5298}}
\author[5]{Pietro Zucca}

\affiliation[1]{National Centre for Radio Astrophysics, Tata Institute of Fundamental Research, S. P. Pune University Campus, Pune 411007, India}
\emailAdd{sohamd943@gmail.com}
\affiliation[2]{Udaipur Solar Observatory, Physical Research Laboratory, Dewali, Badi Road, Udaipur-313 001, Rajasthan, India}
\affiliation[3]{NASA Goddard Space Flight Center, 8800 Greenbelt Road, Greenbelt, MD, 20771, USA}
\affiliation[4]{The Catholic University of America, Washington, DC, USA}
\affiliation[5]{ASTRON – The Netherlands Institute for Radio Astronomy, Oude Hoogeveensedijk 4, 7991 PD Dwingeloo, The Netherlands}

\abstract{Solar radio bursts are generated through the plasma emission mechanism by beams of accelerated electrons travelling along magnetic field lines.
They serve as powerful probes of the ambient coronal plasma and magnetic fields.
At low frequencies, however, propagation through the turbulent, magnetised coronal plasma strongly modifies their observed properties.
Disentangling these propagation effects from the intrinsic nature of the emission therefore remains a major challenge.
Despite the long history of observations, there are only a few that leverage high-angular-resolution imaging.
This is primarily due to the prevailing expectations that scattering from coronal density inhomogeneities significantly increases the apparent source size, obscuring any fine-scale structures that might be present. 
While small-scale features are routinely observed in EUV and X-rays, few are detected at meter wavelengths.
Numerous studies have focused on quantifying the effect of scattering on the observed apparent size of a point solar radio source. 
Naturally, their results vary significantly depending on the specifics of the model used to describe coronal inhomogeneities and the scattering prescription applied -- ranging from sub-arcseconds to arcminutes. 
Recent high angular resolution observations with the Square Kilometre Array Observatory (SKAO) -pathfinders, uGMRT and LOFAR, have revealed unexpectedly compact sources, with sizes as small as $\sim9$ arcseconds in type-I noise storms.
Forthcoming high-fidelity and angular resolution SKAO observations will transform our understanding of solar radio emissions by constraining scattering models, improving coronal diagnostics, and revealing fine structures crucial for understanding solar radio emissions and propagation effects.
}

\begin{document}
\maketitle

% \section{Introduction}
% \begin{itemize}
%     \item Introduction to solar radio bursts
%     \item Need for disentangling propagation effects from intrinsic emissions.
%     \item Challenges in doing so ...
%     \item What high angular resolution observations have to offer to play a role in it.
%     \item Structure of the chapter
% \end{itemize}

\section{Introduction}

The Sun is a prodigious source of radio emission, particularly during episodes of enhanced activity such as flares and eruptive events. Among these, solar radio bursts, transient and often intense emissions observed in meter-decameter wavelengths -- are of special interest because they arise from streams of accelerated electrons propagating through the corona along magnetic field lines and generating coherent plasma emission \citep{Gary_2004,benz2017}. 
These bursts (Type~I--V, spikes, zebra patterns, herringbones, etc.) provide a direct window into particle acceleration, magnetised plasma dynamics, and the ambient conditions of the solar corona and inner heliosphere \citep{ReidRatcliffe2014, chen2018}.

Yet, despite decades of observations, interpreting solar radio bursts in detail remains a challenging task. A key reason is that what we observe is not merely the intrinsic emission from the source region but also the imprint of radio-wave propagation through a strongly inhomogeneous and magnetized plasma. In particular, scattering and refraction by density fluctuations in the turbulent coronal plasma can substantially alter the apparent location, size, brightness, and polarization of the burst \citep{steinberg1971, Kontar19_Scatextn, sharma2020}. 

Disentangling these propagation effects from the underlying emission physics is therefore crucial. On the one hand, the intrinsic radio burst emission encodes information about the acceleration process, the beam-plasma interaction, the nature of the plasma emission mechanism (e.g., fundamental versus harmonic), and the magnetic field geometry. On the other hand, the propagation effects encode information about the coronal density turbulence, anisotropy of fluctuations, and scattering mean-free paths -- thereby probing the ambient coronal medium itself \citep{robinson1983, krupar2018, Kontar19_Scatextn}. 

However, the challenge of separating intrinsic from propagation signatures is formidable. High-angular-resolution imaging of solar radio bursts has historically been limited, particularly at meter wavelengths, where scattering is strongest, and arrays with sufficient baseline coverage have been scarce. This means that many prior studies were unable to determine whether large apparent source sizes reflect intrinsic extended emission or are dominated by scattering and multi-path propagation. 
Conversely, some features usually attributed to propagation may be intrinsic properties of the source that remain poorly characterised.
Observations made close to the source region, such as those by Parker Solar Probe \citep[PSP;][]{Fox2016} during its close encounters, where propagation effects are less pronounced \citep{Jebaraj2023}, offer a way to test this.
We discuss the synergy between such observations and SKA in Section~\ref{sec:space}.
Additionally, modeling of propagation effects involves considerable uncertainties: the spectrum of density fluctuations (inner/outer scales, anisotropy), the radial variation of turbulence amplitude, refraction versus scattering balance, and the geometry of the beam and magnetic flux tube all feed into predictions for observed source morphology and size \citep{kontar2023}. 
Some of these quantities are now being measured in situ by PSP at heliocentric distances ($\approx$10~$R_\odot$ and beyond).
For example, \citet{Krupar2020} measured relative density fluctuations of $\epsilon \approx 0.06$--$0.07$ at 35.7~$R_\odot$, and density fluctuation spectra can be derived from PSP electron densities obtained by quasi-thermal noise spectroscopy \citep[][]{Moncuquet2020}.
Such measurements provide independent constraints on the turbulence parameters assumed in scattering models (Section~\ref{sec:space}).
The emission mechanism itself may also exhibit small-scale structures (in space, time, and frequency) that become smeared out or masked by propagation. Yet such fine structure is crucial for probing the microphysics of beam injection, plasma waves, mode conversion, and emission escape. To isolate such effects, one must have imaging of sufficiently high fidelity and resolution to identify and distinguish features not dominated by scattering or refraction.

High-angular-resolution observations thus offer a path forward. By resolving compact sources, tracking their centroid motions, measuring source sizes, shapes, and polarization evolution across frequency and time, we can begin to separate propagation from emission effects. For instance, surprisingly compact source sizes (of $\sim$9~arcseconds) have recently been reported in Type~I noise-storm observations using SKA pathfinder, the upgraded Giant Metrewave Radio Telescope \citep[uGMRT;][]{Gupta_2017}, suggesting that under certain coronal conditions scattering is less severe than previously assumed \citep{Mondal_2024_gmrt, Mondal2025_typeI_obs_and_modeling}. %, and that we may be directly observing the intrinsic emission \citep{mohan2021, zucca2023, kontar2024}. 
Investigations of source centroid shifts, anisotropic scattering signatures, and simultaneous imaging plus spectroscopy are now feasible. With the arrival of the Square Kilometre Array Observatory (SKAO) and its pathfinders, the combination of high dynamic range, high-angular-resolution imaging, wideband spectroscopy, and polarimetry promises to revolutionise our ability to characterise both solar radio bursts and the coronal propagation medium. In turn, this will enhance diagnostic capabilities for coronal magnetic fields, turbulence levels, and beam propagation paths, ultimately improving our understanding of how particle accelerations in the low corona lead to radio emissions and their escape — a crucial aspect of space-weather physics.

The chapter is structured as follows.  
Section~\ref{sec:propagation} introduces the principal radio--wave propagation effects in the solar corona, including refraction, scattering, and magnetoionic mode coupling, and outlines their theoretical foundations and observable consequences for solar radio imaging.  
We then summarize how these effects manifest in measurable quantities such as apparent source size, centroid displacement, polarization, and brightness temperature, and discuss how they complicate the interpretation of intrinsic emission properties.  
Section~\ref{sec:results_precursors} reviews key results obtained with SKA precursors and pathfinders, highlighting recent high--angular--resolution observations of Type~I noise storms, Type~II bursts, and Type~III bursts that challenge widely accepted scattering models.  
In Section~\ref{sec:ska_advancements}, we describe how the Square Kilometre Array Observatory will overcome longstanding limitations through its high sensitivity on long baselines and extensive frequency coverage.  
Finally, we summarize the implications of these advances for future studies of coronal turbulence, magnetic structure, and particle acceleration in the SKAO era.

By covering both the emission physics and the propagation effects with a forward-looking view towards high-resolution imaging capabilities, this chapter aims to provide a roadmap for achieving a more complete understanding of solar radio bursts in the SKAO era.

\section{Effects of Coronal Radio-Wave Propagation}
\label{sec:propagation}

The radio waves generated as bursts in the solar corona traverse a complex, magnetised and turbulent medium before arriving at ground-based or space-based instruments.  The propagation of these electromagnetic waves through density fluctuations, refractive index gradients and magnetic field structures imposes modifications to the intrinsic emission that must be disentangled in order to realise the full diagnostic potential of solar radio bursts.  In this section, we detail the principal propagation processes and their observable consequences and the implications for high-angular-resolution radio imaging.

\subsection{Key propagation phenomena: refraction, scattering and mode-coupling}

Radio waves in the solar corona obey the dispersion relation for a cold, magnetized plasma. Here we list the key propagation effects encountered by the radio emission:
%Radio waves in the corona encounter several propagation effects:

\begin{itemize}
  \item \textbf{Refraction:} The refractive index $n$ for a wave with angular frequency $\omega$ is given by
  \begin{equation}
      n(r) = \sqrt{1-\frac{\omega^2_{pe}(r)}{\omega^2}}
  \end{equation}
  where $\omega_{pe} = \sqrt{4\pi n_e(r) e^2/m_e}$ is the local electron plasma frequency determined by the electron density $n_e(r)$ \citep{stix1992, melrose1980}.
  
  This relation is central to low-frequency solar radio observations as the dominant meter wavelength bursts, e.g., type I, II and III bursts, are generated via the plasma emission mechanism \citep{melrose1980, benz2017}. 
  This coherent process produces radiation at the fundamental frequency ($\omega \approx\omega_{pe}$) and its harmonic ($\omega \approx 2\omega_{pe}$).
  For fundamental emissions, the refractive index is very low close to the emission site, and the wave is extremely susceptible to density fluctuations \citep{melrose1980, ReidRatcliffe2014}.
    Harmonic emission, generated at $\omega \approx 2\omega_{pe}$ where $n \approx 0.87$, is much less susceptible to these effects near its source (Section~\ref{sec:FvsH}).

  As the refractive index varies with position through its dependence on the coronal electron density, radio waves propagating through the corona experience systematic bending of their trajectories.  
  In the presence of large-scale density gradients, rays are refracted away from regions of higher density, leading to apparent outward shifts of source positions and modifying the mapping between the observed frequency and the true emission height \citep{steinberg1971, arzner1999, Kontar19_Scatextn}.  
  These refractive effects are strongest at low frequencies and are important to account for when interpreting the location and morphology of meter-wavelength radio sources.

  \item \textbf{Scattering:} While refraction is governed by the large-scale coronal density profile, scattering arises from small-scale electron density fluctuations associated with coronal turbulence \citep{robinson1983, bastian1994}.  
  These fluctuations introduce stochastic perturbations to the refractive index, acting as random phase screens that progressively distort the wavefront as radio waves propagate through the corona \citep{arzner1999}.  
  The cumulative effect is angular diffusion and multi-path propagation, leading to angular broadening, temporal smearing, and partial depolarisation of the observed radio emission.  
  At meter wavelengths, scattering is particularly efficient, often dominating the apparent size and morphology of radio sources and producing a strong increase of observed source size toward lower frequencies \citep{bastian1994, Kontar19_Scatextn, zhang2021}.  
  As a result, the spatial scales inferred from low-frequency imaging frequently reflect propagation-induced scattering rather than the intrinsic dimensions of the emission region, with anisotropic turbulence further imprinting direction-dependent source broadening \citep{Kontar19_Scatextn, chen2018}.
    Most of the observational constraints underlying these results come from instruments or observing modes with short baselines, for example LOFAR tied-array imaging, which uses only the core stations with baselines of $\sim2$ km \citep[e.g.][]{2017NatCo...8.1515K}.
  Interferometric studies using LOFAR remote stations, with baselines of tens of km, remain few \citep{Zhang2020, Maguire2021, murphy2021, kumari2025, Morosan2025}.
  Even these may not fully exploit the longest baselines in snapshot imaging: \citet{kumari2025} estimate effective baselines of only $\approx$15~km out of $\approx$85~km available.

  \item \textbf{Mode coupling and depolarisation:} In a magnetised coronal plasma, radio waves propagate in the ordinary (O) and extraordinary (X) magnetoionic modes, which experience different refractive indices and group velocities \citep{stix1992, melrose1980}.  
  Spatial gradients in plasma density and magnetic field strength, particularly in quasi-transverse regions where the wave vector becomes nearly perpendicular to the magnetic field, can lead to partial coupling between these modes during propagation \citep{cohen1960}.  
  Such mode coupling, together with differential refraction and scattering of the O and X modes, can modify the observed polarization state of the radiation, leading to reduced degrees of circular or linear polarization and, in some cases, apparent polarization reversals.  

  Observations made close to the source often indicate that the emission can retain a much higher degree of circular polarization than is typically observed from the Earth.
  During PSP close encounters, fundamental type III emission is circularly polarized at $\approx$50\% on average between 19 and 1.5~MHz, above 90\% in some bursts, much higher than reported from 1~au \citep{Jebaraj2023}, and type III storms show circular polarization fractions of up to $\approx$0.4 \citep{Pulupa2025}.
  These measurements sample greater heights than metric observations, so coordinated SKA and PSP observations of the same events are needed to establish how much of the depolarisation observed from the ground is acquired during propagation.

  The terrestrial ionosphere is a further factor in ground-based solar polarimetry.
  Faraday rotation leaves circular polarization unchanged, so the ionosphere is not expected to reduce the observed degree of circular polarization directly.
  Measurements from outside the ionosphere, such as those planned with LuSEE-Night (Section~\ref{sec:space}), offer such a test to quantify the effects of ionosphere in the polarization properties of solar radio bursts.

\end{itemize}

Together, these effects imply that the observed characteristics of a radio burst (size, position, drift, polarisation, brightness) reflect a convolution of intrinsic emission properties and propagation‐induced distortions.

\begin{table}[t]
\centering
\caption{Propagation effects causing coronal `seeing' in radio observations of active solar emissions.}
\label{tab:coronal_seeing}
\small
\begin{tabular}{@{}l c l l@{}}
\toprule
\textbf{Effect} & \textbf{Freq. Dependence} & \textbf{Impact on Sources} & \textbf{Diagnostic Value} \\
\midrule
Scattering      & $\propto f^{-0.98}$         & Source broadening;     & Probes density fluctuations \\
                &                          & Anisotropic elongation & ($\delta n_e/n_e$) and turbulence \\[4pt]
\addlinespace
Refraction      & $\propto f^{-2}$         & Radial positional shift & Maps large-scale density $n_e(r)$; \\
                &                          & ($\Delta\theta \sim 0.1$--1\,$R_\odot$) & identifies ducts/streamers \\[4pt]
\addlinespace
Mode Coupling   & Critical frequency at QT region & Depolarization; & Constrain magnetic field \\
                &                          & Polarity reversal      & topology and strength \\
\bottomrule
\end{tabular}
\end{table}

\subsection{Observable consequences for solar radio burst imaging}

The propagation effects described above imprint clear signatures on solar radio imaging observations.  
At low frequencies, these effects often dominate the observed properties of radio bursts, complicating the interpretation of intrinsic source characteristics.  
Key observable consequences include:

\begin{enumerate}
  \item \textbf{Apparent source broadening:}  
  The observed angular size of a radio burst is frequently dominated by scattering due to coronal density fluctuations rather than by the intrinsic extent of the emission region.  
  For example, imaging observations with the LOw Frequency ARray \citep[LOFAR;][]{vanhaarlem2013_lofar} at frequencies near 35~MHz report full-width at half-maximum (FWHM) source sizes of order \(18^{\prime} \times 10^{\prime}\) at heliocentric distances of \(\sim 1.75\,R_{\odot}\), consistent with strong angular broadening induced by coronal turbulence rather than a physically extended source \citep{murphy2021}.

  \item \textbf{Centroid displacements and drift:}  
  Large-scale refraction systematically bends ray paths outward as radio waves propagate through the radially decreasing coronal density.  
  As a result, the observed centroid of a radio source is often displaced to larger apparent heights than the true emission site, introducing biases in source localization and density or height diagnostics inferred from the emission frequency \citep{Reid20_typeIIIrev}.  
  In addition, frequency- or time-dependent centroid motion can arise from variations in refractive and scattering conditions along the line of sight, providing indirect diagnostic information about the intervening plasma.

  \item \textbf{Flattened size--frequency relationships:}  
  In the strong-scattering regime, theory predicts a steep dependence of apparent source size on observing frequency, typically \(\theta \propto f^{-2}\) \citep{bastian1994}.  
  However, high-resolution imaging studies at low frequencies sometimes reveal a weaker size--frequency dependence, suggesting departures from simple isotropic scattering \citep[][\(\theta \propto f^{-0.98}\)]{Kontar19_Scatextn}.  
  Such behavior may indicate anisotropic turbulence, saturation of scattering at large heliocentric distances, or favorable viewing geometries that partially suppress angular broadening.

  \item \textbf{Temporal and spectral smearing:}  
  Multi--path propagation through a turbulent coronal plasma leads to a distribution of photon arrival times, producing temporal broadening of impulsive radio bursts and a reduction in spectral coherence.  
  Such scattering--induced delays add to the intrinsic temporal evolution caused by electron beam velocity dispersion, resulting in broadened burst profiles, flattened apparent frequency drift rates, and suppression of rapid temporal fine structure.  
  These effects have been identified in both observational and theoretical studies of Type~III bursts, where low--frequency emission exhibits longer decay times and smoother time--frequency structure than expected from intrinsic beam dynamics alone \citep{arzner1999,2018A&A...614A..69R, Kontar19_Scatextn}.  
  As a consequence, fine substructures such as striae or narrowband burst elements may be partially or completely masked at low radio frequencies.

  \item \textbf{Depolarisation:}  
    The degree of circular polarization observed in meter--wavelength radio bursts is often substantially lower than predicted by intrinsic plasma emission models.  
    While fundamental plasma emission can be intrinsically highly polarized, propagation effects - including scattering, differential refraction of the ordinary (O) and extraordinary (X) magnetoionic modes, and partial mode coupling in quasi--transverse regions - can significantly reduce the net observed polarization \citep{zheleznyakov1963}.  
    Such depolarisation effects are strongest at low frequencies, where propagation paths are longest and magnetoionic effects are most pronounced, and have been widely reported in observations of Type III bursts and noise storms \citep{melrose1980,robinson1983,ReidRatcliffe2014,Rahman20_typeIIIstats}. For a detailed discussion on how the magnetic field and the propagation geometry affect the observed circular polarization, refer to section 3.5 of the chapter on `Coronal Magnetography using Spectropolarimetry with SKA Telescopes' by \citet{Patra01.2026.SKA}.

  \item \textbf{Suppression of brightness temperature:}  

    Another significant consequences of coronal scattering is the severe suppression of the observed brightness temperature, $T_b$, at low radio frequencies.  
    Coherent plasma emission (fundamental or harmonic) produced by electron beams or Langmuir waves is intrinsically extremely bright, with physical brightness temperatures typically estimated to lie in the range
    \begin{equation}
    T_b^{\mathrm{intrinsic}} \sim 10^{11} - 10^{13}\,\mathrm{K},
    \label{eq:Tb_intrinsic}
    \end{equation}
    \citep{ReidRatcliffe2014}.

    However, in the strong--scattering regime that dominates below $\sim$200~MHz, the observed source area, $A_{\mathrm{obs}}$, is determined by the scattering disk rather than the intrinsic emission region.  
    The scattering area increases rapidly with decreasing frequency, approximately following
    \begin{equation}
    A_{\mathrm{obs}} \propto \theta^2 \propto \lambda^4 \propto f^{-4},
    \end{equation}
    or even more steeply ($f^{-4.1}$ to $f^{-4.2}$) when radially aligned anisotropic turbulence is taken into account \citep{bastian1994,Kontar19_Scatextn}.

    Since the observed flux density, $S_\nu$, is conserved during the scattering process, the measured brightness temperature is diluted by the ratio of the intrinsic source area to the observed scattering area:
    \begin{equation}
    T_b^{\mathrm{obs}} = T_b^{\mathrm{intrinsic}}
    \frac{A_{\mathrm{intrinsic}}}{A_{\mathrm{obs}}(f)}
    \propto f^{+4.0 \rightarrow +4.2}.
    \label{eq:Tb_suppression_eq}
    \end{equation}

    This produces the characteristic steep rise of observed brightness temperature with frequency.  
    At frequencies of 30--80~MHz (e.g.\ LOFAR), $T_b^{\mathrm{obs}}$ is typically measured to be $\sim10^{8}$--$10^{9}$~K, three to four orders of magnitude below the intrinsic value, whereas at frequencies above $\sim$1~GHz (e.g.\ VLA, EOVSA) scattering becomes negligible and the measured brightness temperatures approach the intrinsic range of $10^{11}$--$10^{13}$~K.  
    This effect explains why metric and decimetric solar radio bursts often appear comparatively ``cool'' despite their coherent plasma emission origin \citep{ReidRatcliffe2014,2017NatCo...8.1515K}.

\end{enumerate}

\begin{figure}[htbp]
\centering
\includegraphics[trim=0cm 0cm 0cm 1.1cm, clip, width=\linewidth]{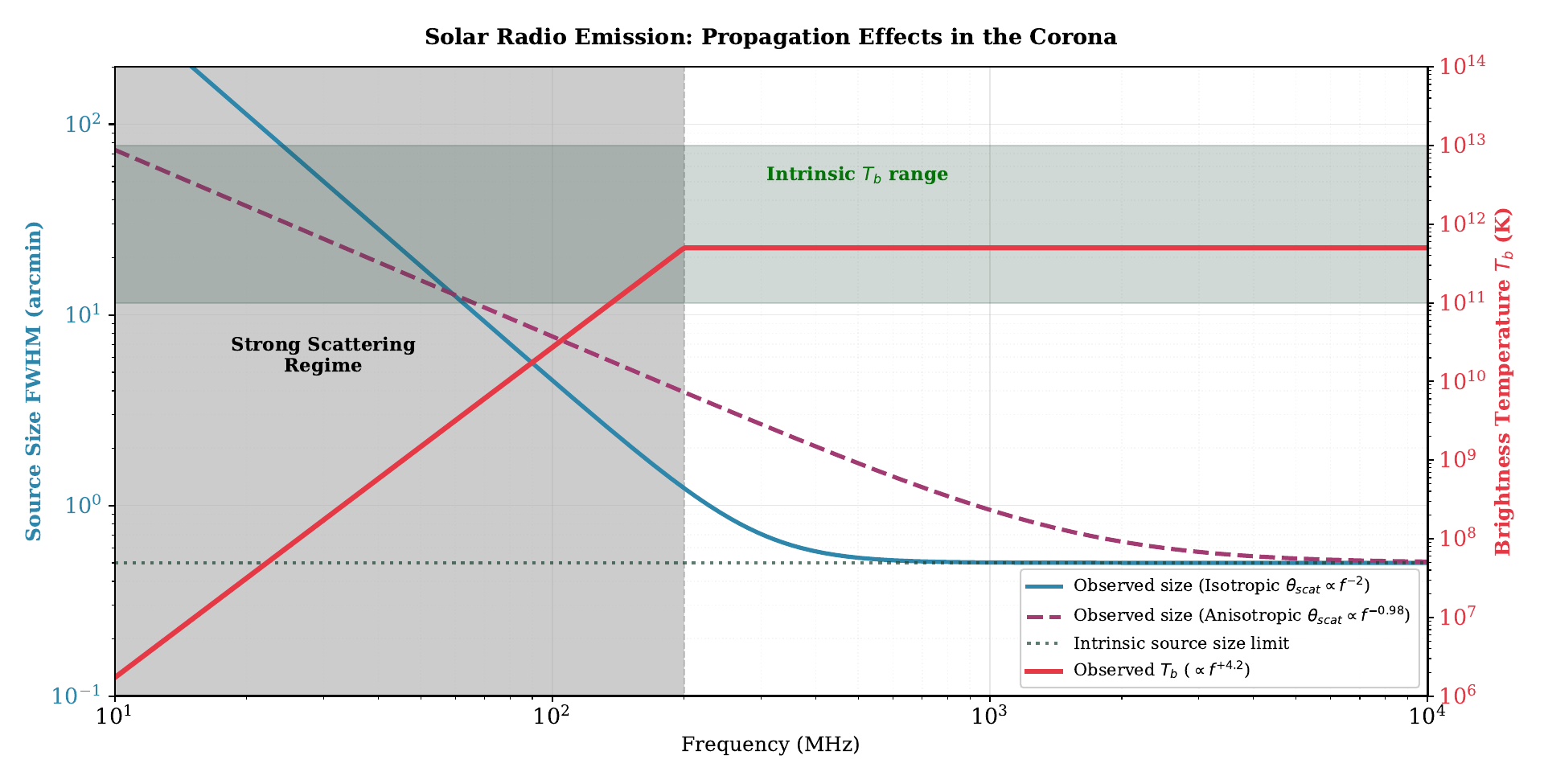}
\caption{Frequency dependence of coronal scattering broadening for solar radio sources.
Solid blue line: isotropic scattering expectation ($\theta \propto \lambda^2 \propto f^{-2}$). 
Dashed magenta line: anisotropic scattering model ($\theta \propto f^{-0.98}$). 
Solid red line: observed brightness temperature.}
\label{fig:figure1}
\end{figure}

\begin{figure}[htbp]
\centering
\includegraphics[width=\linewidth]{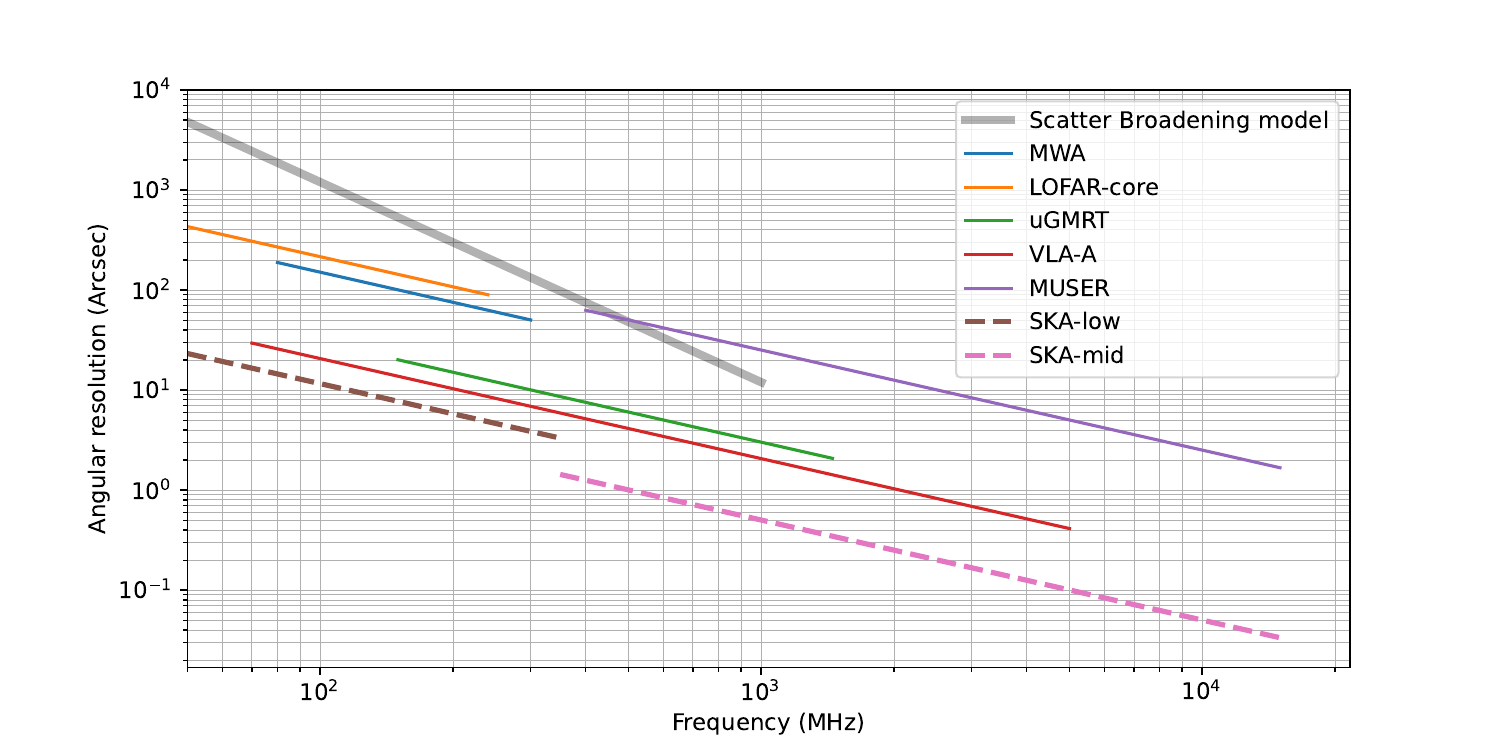}
\caption{Comparison between angular resolution of existing radio interferometers and the SKA telescopes. With their very long baselines, it is expected that the SKA telescopes will be able to observe much smaller sources than currently possible.}

\label{fig:resolution}
\end{figure}

\section{Characterizing Coronal `Seeing' Using SKAO}
\label{sec:ska_defeats_seeing}

The SKAO with its exceptional sensitivity, angular resolution, and wide frequency coverage will make it possible to probe the coronal 'seeing' in both weak and strong scattering regimes. Spectropolarimetric imaging using the SKA telescopes will put constraints on underlying emission mechanisms and particle acceleration sites. This will help putting data driven constraints on the scattering models of radio waves propagating through corona.

The angular size of the scattering disk in the strong anisotropic regime is described by
$\theta_\mathrm{scat}(f) \propto f^{-0.98}$ \citep{Kontar19_Scatextn}.
The synthesized beam of an interferometer scales as
\begin{equation}
\theta_\mathrm{inst}(f) \simeq  
\frac{\lambda}{B_\mathrm{max}},
\label{eq:beam}
\end{equation}
where $B_\mathrm{max}$ is the maximum baseline.

%Figure~\ref{fig:ska_defeats_seeing} compares these two scales:

\begin{itemize}
\item \textbf{SKA1-Low} (50--350 MHz, $B_\mathrm{max} \simeq 65$~km core):  
  $\theta_\mathrm{inst} < \theta_\mathrm{scat}$ already above $\sim$120--130\,MHz — an instrument with unprecedented sensitivity and dense uv-coverage, capable of resolving the \emph{intrinsic} morphology of metric noise storms, Type III beams, and fine structures.

\item \textbf{SKA1-Mid} (0.35--15 GHz, $B_\mathrm{max} \simeq 150$~km):  
  delivers 50--500\,mas resolution, surpassing even the VLA A-configuration and providing the highest-resolution microwave imaging of flaring loops, gyro-resonance layers, and plasma emission sources ever achieved.
\end{itemize}
Figure \ref{fig:resolution} shows a comparison of angular resolution between existing radio interferometers like MWA, LOFAR, uGMRT with the upcoming SKA telescopes. As evident from the plot, SKAO will be able to achieve much higher angular resolution, especially with unprecendented sensitivity at longer baselines, making it possible to study small sources and test existing theories about scattering in corona.

\subsection{Implications for interpreting high-resolution radio imaging}

For high--angular--resolution studies, such as those envisaged for the Square Kilometre Array Observatory (SKAO) era, it is essential to explicitly account for radio--wave propagation effects when interpreting solar radio imaging observations.  
At meter and decameter wavelengths, the observed properties of radio bursts are often dominated by the combined effects of scattering and refraction, and neglecting these processes can lead to systematic misinterpretation of intrinsic source characteristics.

The measured angular size of a solar radio source at any given frequency represents a convolution of the intrinsic emission region with a propagation--induced scattering kernel.  
In the strong--scattering regime, the apparent source size may be set almost entirely by the scattering disk, rather than by the physical extent of the emission region itself.  
Without independent constraints on the scattering properties of the corona, direct inferences of intrinsic source size, emission volume, beam geometry, or brightness temperature can therefore be severely biased, typically leading to overestimation of source size and underestimation of brightness temperature \citep{bastian1994,Kontar19_Scatextn}.

Propagation effects also bias the commonly used mapping between observed emission frequency and coronal height.  
Standard analyses often assume rectilinear propagation and infer source heights by directly associating the emission frequency with a plasma density model.  
However, refraction and scattering systematically bend radio--wave trajectories and displace apparent source centroids outward from their true emission locations, particularly at low frequencies.  
As a result, coronal density and height estimates derived without accounting for propagation effects may be significantly in error, a phenomenon that has been repeatedly identified in imaging studies of Type~III bursts and other low--frequency emissions \citep{steinberg1971,arzner1999,morosan2014}.

Fine spatial and temporal structure intrinsic to the emission process may also be suppressed by propagation through the turbulent corona.  
Sub--arcminute spatial features, rapid temporal variability, and narrowband spectral structures can be partially or fully smeared before reaching the observer, even when instrumental resolution would otherwise be sufficient to resolve them.  
The detectability of such fine structure therefore depends not only on instrumental capabilities but also on the scattering regime along the line of sight, with stronger scattering progressively erasing intrinsic small--scale features \citep{2018A&A...614A..69R,zhang2021}.

Importantly, propagation effects need not be viewed solely as a limitation.  
When interpreted correctly, scattering and refraction themselves become powerful diagnostics of the coronal plasma.  
By measuring the frequency dependence of source sizes, centroid displacements, temporal broadening, and polarization properties, and by combining high--fidelity imaging with spectroscopy and polarimetry, it is possible to constrain key parameters of coronal turbulence, including the density fluctuation amplitude \(C_N^2(r)\), the degree of anisotropy, and characteristic inner and outer scales of the turbulent spectrum \citep{Kontar19_Scatextn}.  
In this way, high--resolution radio imaging enables a dual scientific return, providing access to both the intrinsic physics of solar radio emission and the structure and dynamics of the coronal propagation medium itself.

\subsection{Fundamental versus harmonic emission}
\label{sec:FvsH}

The propagation effects described above act very differently on fundamental (F) and harmonic (H) plasma emission, and the characterisation in this section applies primarily to fundamental emission.
F emission is produced at $\omega \approx \omega_{pe}$, so it is strongly refracted and scattered close to its source.
H emission is produced at $\omega \approx 2\omega_{pe}$, where $n \approx \sqrt{3}/2 \approx 0.87$, and is therefore much less affected in the region where scattering is most efficient.

Recent LOFAR observations using remote baselines support this picture.
For a type IIIb--III pair, the apparent F source at 26.56~MHz expanded from $\approx$50 to $\approx$200~arcmin$^2$ within 0.45~s and moved at an apparent speed of $\approx$4$c$, while the H source at the same frequency remained stable in size and position \citep{Zhang2020}.
For a type II burst, the F sources were displaced radially outward by 0.3--0.5~$R_\odot$ relative to the H sources and were 1.6--1.9 times larger after normalising by the point spread function, while the H positions matched the emission heights expected from a coronal density model \citep{Maguire2021}.
\citet{kumari2025} measured F type II sources of $1^\prime$--$5^\prime$ at 30--70~MHz and H sources of $\approx1^\prime$ at 145--180~MHz and $\approx2^\prime$ at 55--85~MHz, which they report as only slightly exceeding the beam of their effective array.
Imaging of harmonic type II lanes has also resolved herringbone sources of $\approx4^\prime$ FWHM at 45--73~MHz \citep{Zhang2024} and closely spaced but distinct emission regions along the shock \citep{Morosan2025, normo2025}.

The F--H difference is nevertheless not yet well quantified.
Harmonic emission still accumulates scattering along its path, and at a given observing frequency it originates higher in the corona, where $\omega_{pe} = \omega/2$.
At $\approx$32~MHz, \citet{Chen2020} found harmonic type III sources with apparent sizes comparable to fundamental ones, although their temporal evolution was much slower.
%Compact sources are also not restricted to harmonic emission. \citet{Mondal2025_typeI_obs_and_modeling} note that the small noise-storm structures they observe would be even harder to reconcile with scattering if the emission is fundamental.
Simultaneous long-baseline imaging of F--H pairs across the SKA-Low band will allow the differential effect of propagation on the two components to be measured directly.

\section{Results with SKA Precursors and Pathfinders}
\label{sec:results_precursors}

High–angular‐resolution observations of solar radio bursts, although very few are available in the literature, have historically provided crucial insights into their morphology and the limits imposed by coronal scattering.  
\citet{lang1987vla} used the Karl G. Jansky Very Large Array \citep[VLA;][]{Perley2011_vla} at 328 MHz with a spatial resolution of $9^{\prime\prime}$ observing a noise storm source for about three hours. 
They found that the noise storm consisted of multiple compact sources, each of size $\sim40^{\prime\prime}$, located within an extended emission region.  
\citet{Zlobec1992} conducted similar observations at 333 MHz but did not detect correlated flux on baselines longer than 5000$\lambda$ (corresponding to $\sim40^{\prime\prime}$), concluding that the smallest morphological structures inside the noise storm could not be smaller than this limit.  
Subsequent studies by \citet{Mercier2006} and \citet{Mercier2015}, combining data from the legacy Giant Metrewave Radio Telescope \citep[GMRT;][]{Swarup1991} and the Nançay Radio Heliograph \citep[NRH;][]{Kerdraon1997_NRH}, achieved snapshot imaging with angular resolution $\approx20^{\prime\prime}$ at 236–327 MHz.  
They reported that the storm comprised multiple bright ``knots'' embedded within a diffuse halo, with the smallest structures ranging from $31^{\prime\prime}$ to $35^{\prime\prime}$.  
At much lower frequencies, there is even indication of structures as small as $\le15^{\prime\prime}$ at 53 MHz using a two–element interferometer of $\sim200$ km baseline \citep{mugundhan2018}. 
While it is generally believed that coronal scattering effectively smears out small–scale structures, imposing a lower bound on detectable sizes \citep{bastian1994, Kontar19_Scatextn}, these results suggest that
%However, the detection of sub–20" features by \citet{mondal2024, mondal2025} suggests that 
scattering may not always dominate, and that compact intrinsic structures can be observed under favourable coronal or geometric conditions.  
%It has also been proposed that the scarcity of small observed source sizes may stem primarily from the limited angular resolution of instruments historically used for solar radio imaging \citep{Subramanian2011}.  

Building upon these results, the SKA precursors and pathfinders have already dramatically enhanced our ability to probe intrinsic fine structure and to quantify propagation effects.  
In this section, we focus on recent advances achieved using SKA precursors and pathfinder arrays—specifically, the Murchison Widefield Array \citep[MWA;][]{Tingay2013_MWA}, LOFAR, and uGMRT and the new understanding they have yielded of Type I, Type II, and Type III solar radio bursts.

\subsection{Type-I Noise Storms}
\label{sec:typeI_noise}

Type I noise storms are among the most persistent forms of solar radio activity at meter wavelengths, consisting of a quasi–continuous background continuum superimposed with rapid, narrowband bursts \citep{elgaroey1977solar}.  
They typically originate in active regions and are interpreted as signatures of quasi–steady electron acceleration and plasma emission near magnetic loop tops \citep{iwai2012, Mercier2006}.  

\begin{figure}[htbp]
\centering
\includegraphics[width=\linewidth]{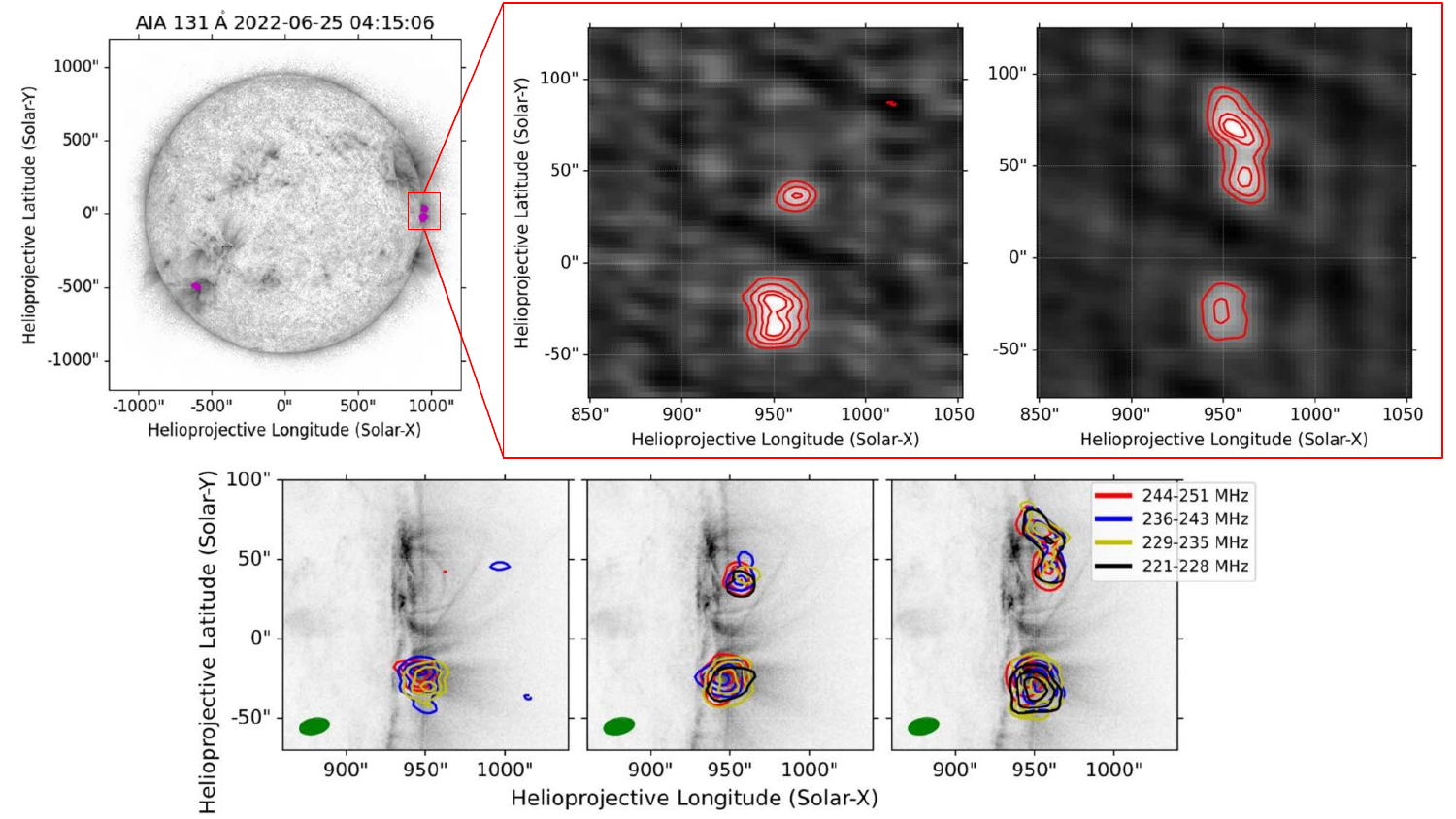}
\caption{
Top left image shows the radio contours overplotted on AIA 131 \AA image. Two sample images from 244-251 MHz band are shown in the zoomed-in images. Some of the smallest structures are seen in this frequency. The bottom panels show evolution of sources in different frequency bands in three different timestamps.
Adapted from \citet{Mondal_2024_gmrt}.}
\label{fig:mondal2024}
\end{figure}

%Recent observations with the upgraded Giant Metrewave Radio Telescope (uGMRT) have achieved the highest angular resolution to date in the 250–500 MHz range, providing new insights into noise–storm morphology and the role of coronal scattering.  
Recently, \citet{Mondal_2024_gmrt} used high-fidelity snapshot imaging using the uGMRT at 220--250~$MHz$ with a resolution of $\approx20^{\prime\prime}\!\times\!9^{\prime\prime}$. Their work revealed that the noise storm emission consists of multiple compact components, some with apparent sizes lower than $\sim9^{\prime\prime}$ (Figure \ref{fig:mondal2024}). They confirmed the presence of compact substructures embedded in diffuse halos, demonstrating that the apparent source morphology evolves dynamically on timescales of less than a minute.  Frequency-dependent motion of centroids and correlated brightness variations between compact components indicate complex spatial and temporal organization within the emission region, likely governed by evolving magnetic connectivity and intermittent electron acceleration.
While the earlier event was located close to the solar limb, in a follow-up study, \citet{Mondal2025_typeI_obs_and_modeling} extended these analyses using wide-band uGMRT observations at 250--500~$MHz$ of several active regions located close to the disc center, and found multiple instances where the source sizes can be smaller than $20^{\prime\prime}$. %The authors demonstrated using simulations of anisotropic scattering that the observed source sizes are significantly more compact under classical scattering prescriptions. 
Together, these uGMRT studies provide the most stringent constraints to date on the lower bound of apparent source sizes in metric noise storms. They demonstrate that imaging with well-calibrated long baselines is capable of resolving sub–20" features. These source sizes are significantly more compact compared with classical scattering prescriptions. 
These findings motivate a re-evaluation of scattering models and suggest that, in some coronal environments, intrinsic structures may dominate observed morphologies even at meter wavelengths \citep{Mondal2025_typeI_obs_and_modeling}.

\subsection{Type-II Bursts}
\label{sec:typeII}

Type II radio bursts arise from shock fronts driven by coronal mass ejections (CMEs) or flare blast waves, where shock–accelerated electron populations excite Langmuir waves that subsequently generate radio emission near the local plasma frequency and its harmonic \citep{cairns2003,kumari2025}.  
Because the emission tracks the motion of a shock through the corona, Type II bursts provide a diagnostic of shock kinematics, coronal density structure, and magnetic field strength.  

Recent observations with the LOw Frequency ARray (LOFAR) have provided new insights into the spatial properties of these bursts. 
\citet{kumari2025} conducted detailed imaging spectroscopy of several Type II bursts in the 10–200 MHz range, utilizing LOFAR's high time-frequency sampling to measure source sizes and centroid motions with arcminute precision.
A key finding of their study is the compactness of the radio sources; measured angular extents of the fundamental and harmonic lanes ere found to be as small as $\sim1.5^\prime$ at 150 MHz.
These sources often exhibit elliptical shapes, likely aligned with the local magnetic field or the orientation of the shock front. 

Complementing these size measurements, \citet{normo2025} and  \citet{zucca2025} utilized LOFAR to investigate the spatial nature of fine spectral substructures, specifically ``band-splitting''. 
While traditional models often attribute split bands to upstream and downstream emission from a co-spatial region, the high angular resolution imaging capability with LOFAR revealed that the split bands originate from spatially distinct regions along the shock front, in these cases.

%The LOFAR observations provide arcminute to sub-arcminute resolution with high time–frequency sampling, enabling direct measurement of source sizes, centroid motions, and fine substructures.  
%They found that Type II sources are often composed of multiple, spatially separated components associated with different sections of the propagating shock. 
%Measured angular extents of the fundamental and harmonic lanes are frequently smaller than 1', indicating that scattering broadening at these frequencies may be less severe than predicted by traditional models.  

\subsection{Type--III bursts}
\label{sec:typeIII}

Type~III radio bursts are produced by beams of suprathermal electrons propagating along open magnetic field lines and generating plasma emission near the local plasma frequency and its harmonic \citep{Reid20_typeIIIrev}.  
Owing to their rapid frequency drift and broad spectral extent, Type~III bursts probe a wide range of coronal heights within a single event and are therefore particularly sensitive to radio--wave propagation effects.

Imaging spectroscopy of Type~III bursts with modern radio arrays has demonstrated that the observed source morphology and centroid locations are often strongly modified by propagation through the structured and turbulent corona.  
Observations with the LOFAR and the MWA at metric and decametric wavelengths have enabled time-resolved imaging of Type~III sources across large frequency ranges, revealing complex apparent trajectories and extended source morphologies \citep[e.g.][]{Chen13_typeIII,morosan2014,Chen18_typeIII_jet,patrick17_MWAQSL,2018NatSR...8.1676C}.

At low frequencies ($\lesssim300$~MHz), type~III sources observed with LOFAR and MWA typically appear broadened to angular scales of order an arcminute and are often well described by two--dimensional Gaussian morphologies.  
In the framework of isotropic scattering models, such source sizes are consistent with strong angular broadening imposed by coronal density fluctuations \citep{steinberg1971,arzner1999,Kontar19_Scatextn}.  
However, departures from simple isotropic behavior are frequently observed, particularly in active regions with complex magnetic topologies or during more energetic events.

Several imaging studies report systematic shifts of Type~III source centroids relative to expectations from standard radial coronal density models, especially for limb events \citep{morosan2014}.  
Such apparent ``overdensities'' inferred from source heights are now widely interpreted as signatures of radio--wave propagation effects - principally refraction, scattering, and ducting - rather than true enhancements in the ambient plasma density.  
Numerical and analytical studies have shown that these effects can displace centroids, alter apparent source trajectories, and bias density diagnostics derived from Type~III imaging \citep[e.g.][]{steinberg1971,robinson1983,arzner1999,Kontar19_Scatextn,Zhang21_SrcSize_dur_pos_turb}.

More recent theoretical work has demonstrated that anisotropic scattering in a magnetised corona can further modify Type~III source morphology.  
In particular, scattering--induced source bifurcation can produce multiple compact components within what would otherwise appear as a single broadened source, leading to apparent source sizes smaller than those predicted by isotropic scattering models \citep{2025ApJ...978...73C}.  
%Consistent with this picture, long--baseline interferometric observations have reported Type~III source sizes as small as $\sim10^{\prime\prime}$ at metric wavelengths, significantly below classical scattering limits \citep[e.g.][]{2016ApJ...831..154M,Mugundhan18_15arcsec_struc}.

Propagation effects also imprint signatures in the frequency--time domain of type~III bursts.  
Multi--path propagation through a turbulent medium introduces additional temporal and spectral broadening beyond that expected from intrinsic beam velocity dispersion alone \citep{1998SoPh..181..363R,1998SoPh..181..395R,arzner1999,Li2009}.  
Disentangling intrinsic electron beam dynamics from propagation - induced effects therefore requires joint analysis of image--plane morphology, centroid evolution, and frequency-time structure \citep{Reid20_typeIIIrev}.

\subsection{Linearly polarized emission at meter wavelengths}

A recent and unexpected result reported by \citet{dey2025} is the robust detection of intrinsic linear polarization in Type~I noise storms and Type~III radio bursts at meter wavelengths using the MWA and uGMRT.  
This finding challenges the long-standing paradigm in solar radiophysics that strong differential Faraday rotation in the solar corona should inevitably depolarize any linearly polarized emission before it reaches the observer \citep{Grognard_McLean_1973, melrose1980}.

A central idea of depolarization, mentioned in Section \ref{sec:propagation}, is that mode coupling in the quasi transverse regions, will lead to lower levels of circular polarization as they are converted to linear polarization which are then depolarized during propagation through the coronal plasma \citep{melrose1974}.
These observations demonstrate that linear polarization can survive coronal propagation under certain conditions.  
The detection of intrinsic linear polarization at meter wavelengths opens a new diagnostic window on coronal plasma conditions and propagation physics.

\begin{figure}[htbp]
\centering
\includegraphics[width=\linewidth]{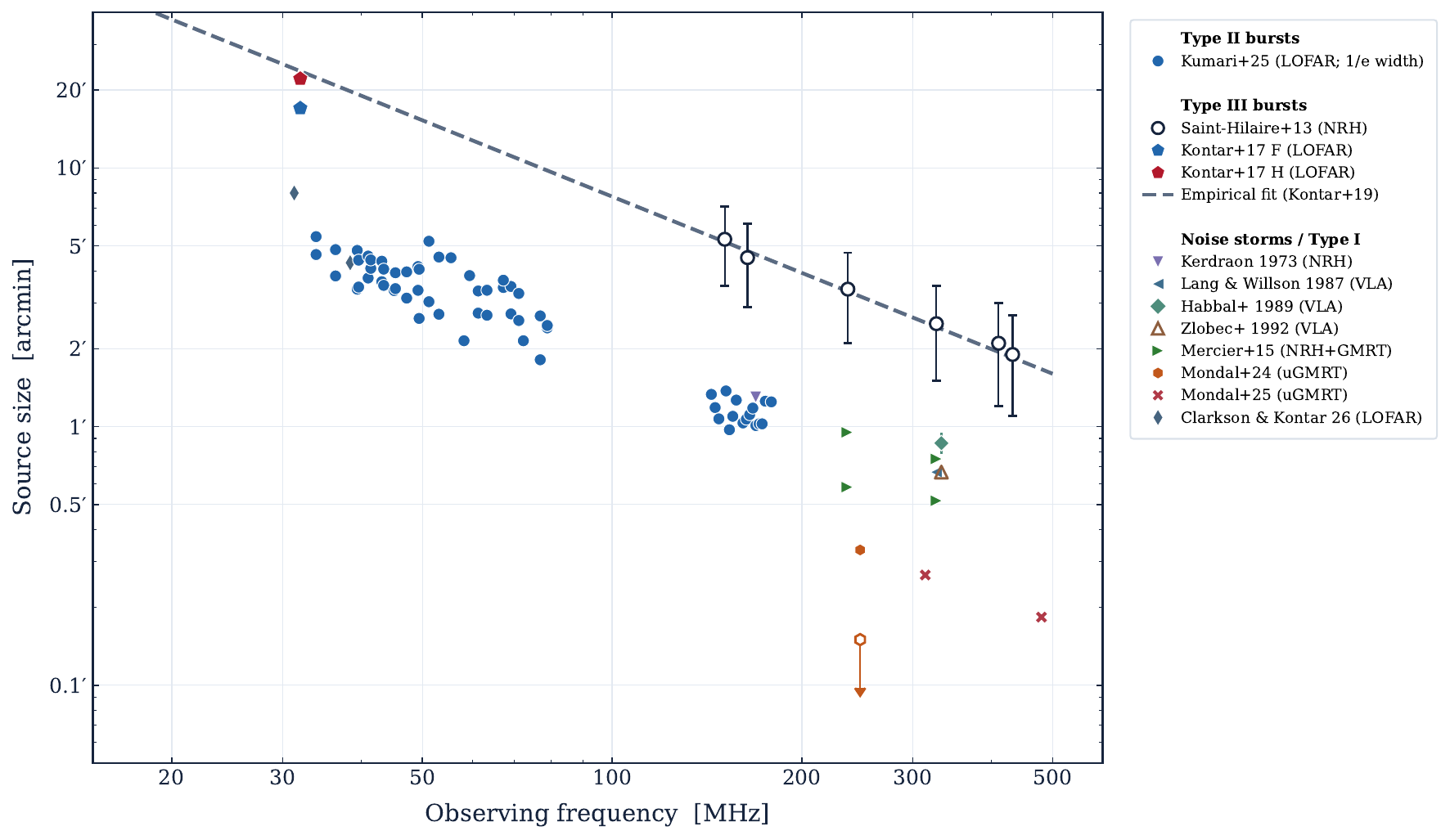}
\caption{Apparent source sizes of solar radio emission versus observing frequency.
Type II bursts: \citet{kumari2025}, LOFAR interferometric imaging; major axes (1/e extent) of 2D Gaussian fits to sources.
Type III bursts: \citet{SaintHilaire2013}, NRH, mean and standard deviation of the deconvolved size $\sqrt{s_a^2+s_b^2}$, where $s_a$ and $s_b$ are the FWHM axes (open circles); \citet{2017NatCo...8.1515K}, LOFAR, beam-corrected sizes $\sqrt{A-A_\mathrm{beam}}$ of F and H sources near 32~MHz, where $A$ is the FWHM area (pentagons).
The dashed line is the fit to type III FWHM sizes compiled between 0.05 and 500~MHz, $\mathrm{FWHM} = 11.78\,f^{-0.98}$~deg with $f$ in MHz \citep{Kontar19_Scatextn}.
Noise storms: sizes reported by \citet{Kerdraon1973}, \citet{lang1987vla}, \citet{Habbal1989} and \citet{Zlobec1992}, as quoted by \citet{Mercier2015}; for \citet{Habbal1989} the marker is the geometric mean of the reported $57^{\prime\prime}\times47^{\prime\prime}$ and the bar spans the two axes. Smallest core minor axes (half-power widths) from \citet{Mercier2015}; structures of $\sim$9$^{\prime\prime}$--20$^{\prime\prime}$ from \citet{Mondal_2024_gmrt}, the $\sim$9$^{\prime\prime}$ value being an upper limit (arrow); deconvolved minor axes from \citet{Mondal2025_typeI_obs_and_modeling}; beam-corrected FWHM major axes from \citet{ClarksonKontar2026}.
All points are plotted at the observing frequency. Size definitions differ between studies, as listed, so the comparison is indicative.}
\label{fig:source_sizes}
\end{figure}

\section{Advancements with the SKA}
\label{sec:ska_advancements}

While substantial progress in understanding coronal radio--wave propagation has been achieved with SKA precursors and pathfinders, the Square Kilometre Array Observatory (SKAO) enables a major leap in this science.  
Two key technical advances are particularly transformative for studies of coronal seeing effects: (i) high sensitivity on long baselines and (ii) continuous, wide frequency coverage spanning both the strong- and weak-scattering regimes.

The SKA--Low and SKA--Mid arrays provide unprecedented sensitivity, angular resolution, and spectropolarimetric capability across a broad frequency range, enabling propagation effects to be quantified.

\subsection{High sensitivity at long baselines}

A fundamental limitation of current low-frequency solar imaging is that arrays with long baselines, while essential for resolving compact intrinsic structure, are often sensitivity-limited. 
As a result, the absence of correlated flux on long baselines in many of the previous observations, could be due to the low sensitivity at long baselines.

The SKA overcomes this limitation through its combination of large collecting area and dense baseline distribution extending to tens of kilometres.  
In the SKA--Low array, baselines extending to $\sim$74~km provide arcsecond--scale resolution across 50--350~MHz, while SKA--Mid baselines extend to $\sim$36~km (and up to $\sim$108~km with the long--baseline outlier), delivering sub--arcsecond resolution at decimetric and centimetric wavelengths.

This dramatic increase in sensitivity on long baselines enables:
\begin{itemize}
    \item Direct detection of compact intrinsic emission components embedded within scattering halos, if they exist
    \item Robust separation of intrinsic source structure from propagation--induced broadening,
    \item Measurement of anisotropic scattering through resolved source elongation and baseline--dependent visibility attenuation,
    \item Tracking of frequency--dependent centroid shifts with high astrometric precision.
\end{itemize}

Crucially, SKA imaging will allow the scattering kernel itself to be constrained observationally, rather than assumed, by comparing correlated flux densities across a wide range of baseline lengths and orientations.  
This capability is essential for testing anisotropic and magnetically aligned turbulence models and for identifying regimes where intrinsic source sizes dominate over scattering.

\subsection{Extensive frequency coverage}

Equally transformative is the SKA’s continuous frequency coverage from tens of megahertz to several gigahertz.  
SKA--Low covers the 50--350~MHz range, where scattering, refraction, and mode coupling are strongest, while SKA--Mid extends observations into the weak--scattering regime above $\sim$1~GHz, where intrinsic source properties can be directly recovered.

This wide frequency leverage enables propagation effects to be studied in a differential and self--consistent manner.  
In particular, SKA observations will allow:
\begin{itemize}
    \item Direct measurement of the frequency scaling of source size, centroid displacement, and brightness temperature across multiple decades in frequency,
    \item Identification of transitions between strong and weak scattering regimes, for broadband sources,
    \item Quantification of frequency--dependent depolarization and mode coupling,
    \item Simultaneous imaging of the fundamental and harmonic components of the same bursts, to quantify the differential effect of propagation on the two.
\end{itemize}

The availability of high time-- and frequency--resolution full Stokes imaging spectroscopy across this full bandwidth further enables propagation effects in the image plane to be linked directly with temporal and spectral smearing observed in dynamic spectra.  
By tracing the same burst across frequencies that sample vastly different coronal heights and scattering conditions, SKA observations will allow coronal density fluctuation amplitudes, anisotropy scales, and characteristic inhomogeneity lengths to be constrained with unprecedented fidelity.

\subsection{Synergies with space-based radio observatories}
\label{sec:space}

Ground-based arrays see solar radio bursts only after the emission has crossed the overlying corona and the terrestrial ionosphere.
Spacecraft receivers offer a complementary view from above the ionosphere and, in the case of PSP and Solar Orbiter, from close to the Sun.
Table~\ref{tab:space_radio} lists the space-based instruments currently observing solar radio bursts; earlier missions are summarised in the chapters by \citet{Kumari01.2026.SKA} and \citet{Morosan01.2026.SKA} in this collection.
Together they observe from heliocentric distances between 9.86~$R_\odot$ and 1~au and from widely separated directions, but all operate below $\approx$20~MHz, below the SKA-Low band.

\begin{table}[t]
\centering
\begin{threeparttable}
\caption{\textbf{Space-based radio instruments currently observing solar radio bursts. Frequency ranges are those of the radio receivers listed. DF: direction finding.}}
\label{tab:space_radio}
\footnotesize
\setlength{\tabcolsep}{3pt}
\renewcommand{\arraystretch}{1.15}
\begin{tabularx}{\linewidth}{@{}l l l >{\raggedright\arraybackslash}X >{\raggedright\arraybackslash}X c@{}}
\toprule
\textbf{Mission / instrument} & \textbf{Receiver} & \textbf{Freq.\ (MHz)} & \textbf{Location} & \textbf{Capability} & \textbf{Launch} \\
\midrule
\multirow{3}{*}{Wind / WAVES\tnote{a}} & TNR  & 0.004--0.256  & \multirow{3}{=}{Sun--Earth L1} & \multirow{3}{=}{DF by spin demodulation (RAD1)} & \multirow{3}{*}{1994} \\
                                      & RAD1 & 0.02--1.04    & & & \\
                                      & RAD2 & 1.075--13.825 & & & \\
\midrule
\multirow{2}{*}{STEREO-A / WAVES\tnote{b}} & LFR & 0.0025--0.16  & \multirow{2}{=}{$\approx$0.97~au, $\approx$70$^\circ$ ahead of Earth\tnote{e}} & \multirow{2}{=}{DF and polarization} & \multirow{2}{*}{2006} \\
                                           & HFR & 0.125--16.025 & & & \\
\midrule
\multirow{2}{*}{PSP / FIELDS-RFS\tnote{c}} & LFR & 0.01--1.7 & \multirow{2}{=}{9.86~$R_\odot$ to 0.73~au} & \multirow{2}{=}{Circular polarization} & \multirow{2}{*}{2018} \\
                                           & HFR & 1.3--19.2 & & & \\
\midrule
\multirow{2}{*}{Solar Orbiter / RPW\tnote{d}} & TNR & 0.004--1  & \multirow{2}{=}{Perihelion $\approx$0.29~au\tnote{f}} & \multirow{2}{=}{DF and polarization} & \multirow{2}{*}{2020} \\
                                              & HFR & 0.375--16 & & & \\
\bottomrule
\end{tabularx}
\begin{tablenotes}[flushleft]
\footnotesize
\item[a] \citet{Bougeret1995}.
\item[b] \citet{Bougeret2008}; DF and polarization by goniopolarimetry. Contact with STEREO-B was lost in 2014.
\item[c] \citet{Pulupa2017}; circular polarization (Stokes~V) from auto- and cross-spectra of two dipoles \citep{Pulupa2025}.
\item[d] \citet{Maksimovic2020}; DF and polarization by goniopolarimetry with the TNR \citep{Vecchio2021}.
\item[e] the separation from Earth changes with time.
\item[f] Orbit inclination $\approx$17$^\circ$, rising to $\approx$24$^\circ$ after the December 2026 Venus flyby.
\end{tablenotes}
\end{threeparttable}
\end{table}

PSP \citep{Fox2016} and Solar Orbiter \citep{Muller2020} observe bursts from close to their source regions, where the propagation path is short and propagation effects are less pronounced.
PSP has shown that fundamental--harmonic (F--H) pairs make up most type III bursts seen during its close encounters, with the fundamental strongly circularly polarized ($\approx$50\% on average between 19 and 1.5~MHz, above 90\% in some bursts) and the harmonic weakly polarized ($\lesssim$30\%) \citep{Jebaraj2023}.
PSP has also observed type III storms that are highly circularly polarized, with fractional polarization up to $\approx$0.4 \citep{Pulupa2025}.
Such near-source measurements help separate intrinsic emission properties from propagation effects.

Observing the same burst from several vantage points adds information that no single instrument provides.
Simultaneous observations of type III bursts by Wind, STEREO, PSP and Solar Orbiter have been used to estimate source positions and the directivity of the emission \citep{Musset2021, Chen2023}.
The directivity varies from event to event and is consistent with anisotropic scattering \citep{Musset2021}, and the apparent sources lie at larger heliocentric distances than the regions where the emission is produced \citep{Chen2023}.
Interplanetary bursts are therefore subject to the same `seeing' as the coronal bursts discussed in this chapter.
SKA-Low images of the same bursts at 50--350~MHz, closer to their origin, will allow apparent source positions to be followed from the low corona into interplanetary space, and the cumulative effect of scattering along the path to be compared with propagation models.

\iffalse
\textbf{PSP also samples the solar-wind plasma in situ.
Its electron density measurements, including those from quasi-thermal noise spectroscopy \citep{Moncuquet2020}, provide the density fluctuation level and spectrum that scattering models otherwise have to assume.
\citet{Krupar2020} measured relative density fluctuations of $\epsilon \approx 0.06$--$0.07$ in situ at 35.7~$R_\odot$, in agreement with values inferred remotely from radio bursts emitted at that distance.
Cross-calibration of SKA scatter-broadening measurements against PSP in situ density spectra is discussed in the chapter `Probing the solar corona and the solar wind using angular broadening observations with the SKA' by Zhang et~al.\ in this volume.}
\fi

SKA-Low ($\geq$50~MHz) and the receivers in Table~\ref{tab:space_radio} ($\leq$19.2~MHz) do not overlap in frequency, but they sample the same events at different heights.
A type III electron beam or a type II shock can be imaged by SKA-Low in the low corona and followed by the spacecraft further out.
\citet{Morosan2025} give an example: type II lanes recorded by PSP at 10--20~MHz, at about half the frequency of those imaged by LOFAR at 30--80~MHz, indicated that the LOFAR lanes were largely harmonic emission.
Ground-based arrays such as LOFAR, whose low band extends to 10~MHz, can bridge the remaining frequency gap.

Observations from outside the ionosphere are particularly valuable for polarimetry.
The Lunar Surface Electromagnetics Experiment--Night \citep[LuSEE-Night;][]{Bale2023}, to be delivered to the lunar far side by Firefly Aerospace's Blue Ghost Mission~2, will make full-Stokes spectral measurements between 0.1 and 50~MHz, a band in which the Sun is expected to be among the brightest sources.
Its band lies just below that of SKA-Low and overlaps the LOFAR low band (10--90~MHz), so coordinated observations of the same bursts would allow polarization measured with and without the intervening ionosphere to be compared directly.
The Sun Radio Interferometer Space Experiment (SunRISE), a six-spacecraft interferometer to operate slightly above geosynchronous orbit, will localize type II and type III bursts at 0.1--23~MHz \citep{Kasper2022}; it is awaiting launch.

\section{Summary}

The investigation of coronal radio-wave propagation, or coronal ``seeing'', is entering a transformative phase. 
At meter and decameter wavelengths, propagation effects such as refraction, scattering, and mode coupling fundamentally shape the observed morphology, position, polarization, and brightness temperature of solar radio bursts, often obscuring their intrinsic properties.  
Historically, these effects imposed severe limitations on solar radio imaging, leading to the widespread assumption that apparent source sizes and locations were dominated by isotropic scattering.

Recent high--angular--resolution observations with SKA pathfinders and precursors have begun to overturn this paradigm.  
Imaging studies of Type~I noise storms, Type~II bursts, and Type~III bursts with instruments such as LOFAR, MWA, and uGMRT demonstrate that compact sub--arcminute, and in some cases sub--20$^{\prime\prime}$, structures can be observed even at metric wavelengths.  
These results indicate that classical isotropic scattering prescriptions are not universally applicable and that anisotropic turbulence, structured magnetic environments, and favorable propagation geometries can allow intrinsic source properties to remain partially visible.
Propagation effects act most strongly on fundamental emission; harmonic sources appear considerably less affected, a difference that SKA can quantify.

The observable consequences of coronal seeing, including angular broadening, centroid displacement, temporal and spectral smearing, depolarisation, and brightness temperature suppression, are now increasingly well characterized.  
Rather than representing unavoidable observational limitations, these effects can be exploited as diagnostics of coronal density fluctuations, turbulence anisotropy, and magnetic field topology when interpreted with appropriate propagation models.

The Square Kilometre Array Observatory represents a decisive step forward in this context.  
Its combination of high sensitivity on long baselines, wide and continuous frequency coverage, and full--Stokes imaging spectroscopy will enable coronal propagation effects to be measured directly and disentangled from intrinsic emission properties.  
By transforming coronal seeing from a limiting nuisance into a quantitative probe, the SKA will allow intrinsic solar radio source physics and coronal plasma structure to be investigated with unprecedented fidelity.

\section{Acknowledgements}

S.D., D.P., D.O. and S.M. acknowledge support from the Department of Atomic Energy, Government of India, under project 12-R\&D-TFR-5.02-0700.

\bibliographystyle{abbrvnat-maxbibnames4}
\bibliography{chapter} % if your bibtex file is called example.bib

\end{document}